# Development of a Virtual Reality Application for Oculomotor Examination Education Based on Student-Centered Pedagogy

Austin Finlayson
*School of Computer Science*
*East Carolina University*
Greenville, USA
finlaysona20@students.ecu.edu

Rui Wu
*Department of Computer Science*
*East Carolina University*
Greenville, USA
WUR18@ecu.edu

Chia-Cheng Lin
*Department of Physical Therapy*
*East Carolina University*
Greenville, USA
*linch14@ecu.edu*

Brian Sylcott
*Department of Engineering*
*East Carolina University*
Greenville, USA
sylcottb15@ecu.edu

*Abstract*— This work-in-progress paper describes how to utilize student-centered pedagogy to teach the concept of clinical oculomotor examination using Virtual Reality (VR) techniques. Oculomotor examination is an important bioinformatics concept which is typically taught with traditional methods such as PowerPoint slides and lab activities. Though these methods may be effective in part, there is an eventual need to provide hands-on experience to students. Providing the necessary equipment to give students the opportunity to conduct clinical oculomotor examinations can be very difficult for many institutions. The oculomotor examination equipment used in clinics by healthcare professionals is expensive, and many courses teaching this topic cannot afford the cost of clinic-level hands-on experience for students. Besides this, healthcare professional students can have different backgrounds, and their learning preferences and pace may vary. To tackle these challenges, a VR-based oculomotor examination application is developed according to student-centered pedagogy, providing a much lower-cost option for a hands-on oculomotor examination experience and offering students the freedom to walk through the oculomotor examination at different paces for personal study. In this project, a VR application is built using the Unity real-time application engine paired with the HTC Vive Pro headset's built-in accelerometer sensors and eye-tracking functions. Sample data and result images can be generated using our VR application, allowing students the opportunity to collect and analyze data directly and view how data has trended over multiple tests. Based on the results of this user study from students in the Doctor of Physical Therapy program, our proposed VR application is promising as a study tool for oculomotor examination. The results show students high preference for flexibility when using our VR app as an educational tool, and this paper also discusses the possible impacts of this in engineering and computing education.

*Keywords – Clinical Training, Oculomotor Examination, VR*

## I. INTRODUCTION

This paper examines the results of a study conducted to measure the efficacy of a student-centered technique for healthcare professional education utilizing virtual reality (VR). This study was designed to compare the learning progress and opinions of students taught oculomotor examination through both virtual reality based on student-centered pedagogy and traditional instructor-centered teaching methods. In many classrooms, these traditional methods consist of slide-based lectures and lab activities to detail the process to students. These methods may not be ideal, as they offer little interaction with students learning, and compel students to follow and accept a single learning path. A better learning environment would provide students with a flexible pace (i.e., student-centered) [1]. Providing students with clinic-level oculomotor examination devices would also be ideal. This would allow students not only to have the ability to handle the equipment they will use as a professional, but also provide access to real data collected by the students. In clinical settings, these devices are very expensive and may not be affordable for many institutions, so students have limited access to these devices. On the other hand, virtual reality devices, such as the HTC Vive headset utilized for this study, provide a cost-effective solution and offer all the features needed to develop an alternative examination application for the classroom.

The efficacy of VR learning in STEM education has been a subject of study for some time [2], as high-fidelity VR has become increasingly affordable. The potential for increased learning outcomes and engagement made possible by the immersive tool seems high. Many fields may be able to benefit from creating a curriculum that utilizes these technologies in the classroom, and there has been a push in the scientific, engineering, and medical fields to make VR learning a reality. This technology is not only researched for the potential increase in student engagement and learning outcomes [3-4], but also for the potential of the included sensors within the device [5], [6]. This paper details an application created to take advantage of the VR headset and its built-in eye-tracking sensors, allowing the collection of oculomotor examination data during the application. This allows an experience near to that of the clinic-level devices for both subject and administrator and allows students to examine collected data from each test directly within the application from the desktop view.

In this paper, our major contributions are to propose this VR application based on student-centered pedagogy [1] for the oculomotor examination training. The potential for this alternative is especially important for institutions without access to clinic-level devices in the classroom, as the relatively low cost of virtual reality devices makes them a viable candidate to replace video and slide-based teaching methods. Additionally, the student-centered pedagogy provides students more freedom to learn the required knowledge and this can enhance students' engagement and promote students' learning experience [1]. The results and analysis of this user study, conducted to evaluate student learning outcomes and opinions, makes evident the potential of student-centered learning frameworks and the improvements to learning outcomes.

In the remainder of this paper, section II details the related studies of VR in education and summarizes their findings. Section III describes the proposed method of the virtual reality application design, data collection, and interface. The details of the user study are also provided. Section IV comprises the discussion of user study results, detailing student responses and feedback. Section V concludes this paper by describing the implications of our results, and the potential impacts on future work in applying VR for education.

## II. RELATED WORK

In clinical settings, computerized oculomotor examination provides an object measurement when compared with observational oculomotor examination. In the past, a computerized oculomotor device required a higher speed eye tracking camera (> 500 Hz) to obtain validated data. However, newer developed computing algorithms make lower speed eye tracking systems (~ 60 to 120 Hz) capable to provide reliable data [7]. A VR headset with a built-in eye-tracking system may be an alternative to help clinicians with obtaining oculomotor function data in the initial screening for potential neurological disorders. We have developed a VR app that can measure basic oculomotor functions, such as saccade and vestibular-oculomotor reflex, based on existing oculomotor examination methods [8-9]. Healthcare students may use this app to learn the concept of a computerized oculomotor examination.

## III. PROPOSED METHOD

### A. Clinical Knowledge and Activities Design

The results of the clinical oculomotor examination can help clinicians with screening a potential neurological disorder. However, the reading of the oculomotor examination is varied based on the clinician's experience. With the development of computerized oculomotor examination, the program can identify normal vs. abnormal oculomotor function. Patients with abnormal oculomotor function may be referred to a specialist, such as a neurologist or an ophthalmologist, for further examination.

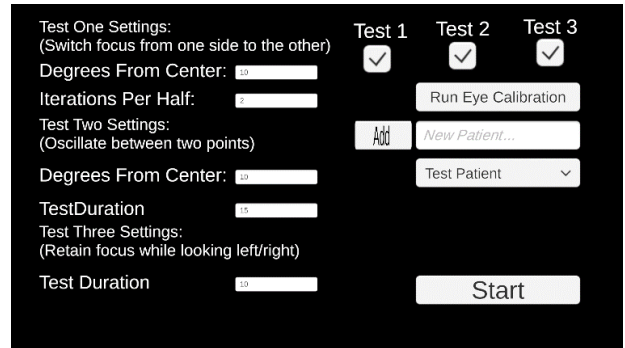

Fig. 1. Application configuration menu. Tests can be selected and configured before start, test variables configured, and a patient can be selected for data logging purposes.

The users who participated in this study completed three eye-tracking tests, designed to allow the user to experience the patient's view during a clinical oculomotor exam [8-9]. The first test consists of three "Focus Objects", which are objects that detect if the user is looking at them. For this test, the user sees a green sphere directly in front of them. Looking at the sphere will cause a red sphere to appear at random to either the left or right of the green one. The user may direct their vision to the center green object to cause the next object to become active after a random period of time between 2 and 5 seconds. The random period can eliminate the prediction of object appearances to obtain optimal results. This test measures the time between a red sphere appearing and the user directing their vision toward it, saving this time at specific intervals of the test as "Latency". The second test shows a red sphere appearing in front of the user and moving back and forth horizontally within the user's field of view. This test measures the users' ability to maintain precise focus on a moving object, saving a measure of focus precision as a value. The third test requires the user to direct their focus to an object fixed in space. The user should rotate their head to the left and right quickly while maintaining focus on the target Focus Object. The number of times the user completes a rotation from left to right is saved as "Frequency", while the speed of the user's head rotation is saved as "Speed". This tests the users' ability to maintain focus on a still object while moving.

### B. Software Implementation

This subsection is about the application implementation, which is available on GitHub. Our application has three components: Eye Data Collection, Visualization, and Log System. Figure 1 shows the dashboard for configuration settings. In this dashboard, which is the initial screen the application opens to, users can select the tests they wish to take, adjust relevant variables for each test (angle between focus objects or angle of travel in tests 1 and 2, length of time for each test in tests 1-3, etc.). Users may also select a patient from the saved profiles in the application or create a new profile. These profiles exist so that the log system may save data for each patient.

The collection of data is performed utilizing the Vive SRAnipal API, developed by HTC. The API can be installed into Unity, allowing eye data to be programmatically accessed by the application during runtime. The API utilizes the eye-tracking sensor built-in to the VR device and uses the sensor to calculate data points such as individual pupil position, gaze direction, gaze origin, pupil diameter, eye openness, and other such variables. The gaze origin and gaze direction vectors, combined with the position vector of a target object, provide all that is needed to detect if the user's gaze collides with that object in virtual space. The gaze origin and gaze direction of each eye are recorded in this application, along with other test-specific variables calculated using them. Visualizations of test data are generated within the application through Python scripts, which use MatPlotLib for graph generation (Figure 2). Data is saved per user so that each user's data trends can be viewed.

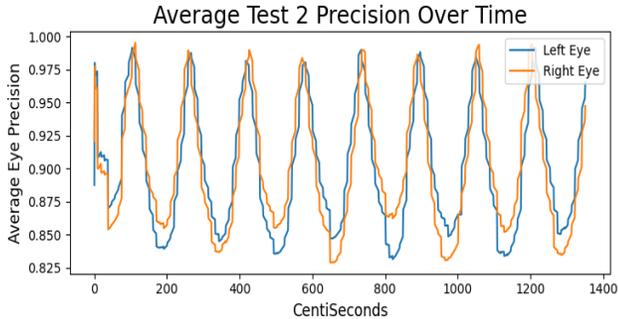

Fig. 2. This line chart shows the precision of the left and right eye over time in test 2. This test has the user track an object which moves to the left and right. The variability in the data is caused by the users' vision falling behind as the object moves across their vision, and catching up as the object slows, stops, and moves the other direction.

This application was developed using the Unity Real Time Development Platform. This platform supports the creation of custom scripts written in the C# programming language. These scripts can be used in traditional ways, such as to process and output data, but can also edit the values of other components within Unity. The development of this application consisted primarily of the creation of custom C# scripts to modify the 3D positions of objects on the VR screen and record data, allowing for the creation of three tests with elements which can interact with eye-tracking data. In test 1, objects alternate to the users' left and right as they are focused on, measuring focus latency (Figure 3). Test 2 moves a singular red sphere left and right across the VR screen at a specified angle as selected in the configuration menu, measuring focus precision. Test 3 has the user rotate their head left and right while maintaining focus on a still object, measuring the users' movement frequency, speed, and gaze precision.

*C. Student-centered Pedagogy*

In a traditional instructor-centered classroom, topics are defined before class and an instructor controls the education flow. This strategy has some challenges. For example, a student might have their own needs or preferences for choosing a learning path and pace, which is difficult to address in a traditional classroom. To solve this issue, a student-centered pedagogy [1] is leveraged in our VR application.

In student-centered pedagogy, students are assisted in controlling their flow of education by using teaching assistants and technologies. Instructors need to ensure students make progress in every class and collect questions from students. Students can work on different topics in the same classroom at different paces.

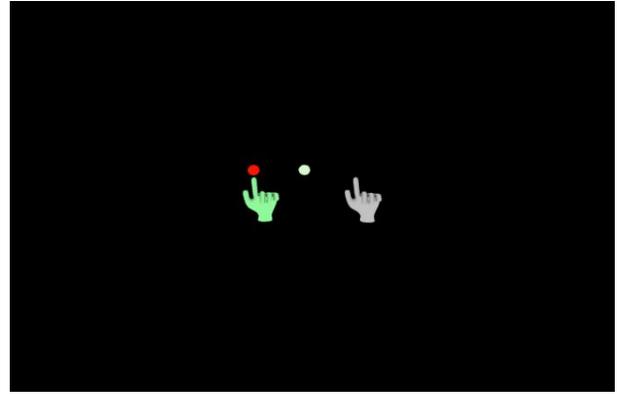

Fig. 3. Test 1 of the application. There are 5 separate objects in the above scene: 3 focus objects (spheres, one is not rendered) and 2 hand objects.

A topic can include multiple components, visualized using a dependency graph (Fig. 4); the vertices represent course content and the edges represent dependencies. The dependencies for both topics and topic components should also be defined but can also be flexible to allow for multiple pathways through the topic; for each student, the order course content is explored can be different. For example, a student should learn basic chemistry principles before understanding mineral composition and physical properties. The dependencies between each course topic and its components will be expressed in the dependency graph structure (Figure. 4). Within our VR application we have three tests. Students should be free to walk through these tests in any order based on their preferences and needs. As this prototype is further developed with the addition of many more tests, the student-centered pedagogy can become more fully realized with additional content and dependencies to be studied at students' own pace.

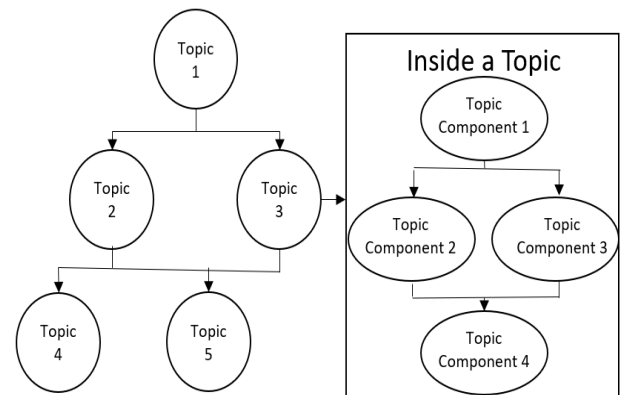

Fig. 4. This figure shows the student-centered learning pedagogy structure. On the left are topics (vertices) structured by dependencies (edges). The right shows the structure within a topic.

## IV. Experimental results and discussion

### A. Experimental Setup

To conduct the user study, 21 Doctor of physical therapy students were first asked to complete a pre-questionnaire before testing. This questionnaire asked students what method of instruction they preferred to be used in various scenarios, with choices of PowerPoint presentations (i.e., instructor-centered), virtual reality (i.e., student-centered), or zero preference between the two. After collecting the students' pre-questionnaire results, students would complete all three tests while wearing the VR headset. A post-questionnaire is then completed after these tests, to gauge whether the student has a new perspective on if VR can be used as a tool for healthcare education, if students are optimistic about this method being used for student-centered learning, and to see if students feel more prepared to conduct an oculomotor examination.

### B. Preliminary Results and Discussion

Twenty-one students in the Doctor of physical therapy program were recruited in this preliminary analysis. From the pre-questionnaire, 57% of physical therapy students on average responded that they preferred PowerPoint slides, with an average of 32% having no preference. 11% of students on average indicated a preference for VR learning. Feedback from users in the post-questionnaire was largely positive, while feedback in key areas was divided. Figure 5 shows user feedback from two of the questions asked on the post-questionnaire. Based on this feedback, it seems many users felt the application offered some insight into how VR could be used as an educational tool, and indicated students felt the application was a promising tool for personal study. In all questions regarding the potential of VR to educate, feedback was overall positive. However, only 43% of users felt that using the application made them more confident to conduct an oculomotor examination, while 28% disagreed and 29% neither agreed nor disagreed. Additionally, 29% of users felt the application made them feel closer to a licensed clinician, while 33% disagreed and 38% neither agreed nor disagreed. It seems that the application was a successful demo for VR as an educational tool in the student-centered pedagogy framework, but the effectiveness as a tool to teach the oculomotor examination was largely split, with more users not feeling an increase in preparedness after using the application.

Based on the feedback received, a likely problem with the application is its perspective; when a user runs through the test, they are not conducting an oculomotor exam, they are taking one. Though some users still felt this made them more confident to conduct an examination, the lack of a clinician's perspective in the application may be why more users did not feel more confident. Two approaches could be made to resolve the issue: the application could be restructured to have the user conduct an exam instead of taking one, or the application could be restructured to feature two users – one taking the exam, and another conducting it. Both solutions could use motion controllers packed in with VR headsets. This would allow the user to interact with the testing environment, allowing greater immersion. This could make significant improvements to how students learn with the application, as evidence suggests that increased interactivity and immersion may lead to better learning outcomes [10-11].

The promising responses to VR as a tool for study indicate the potential for a student-centered learning pedagogy which can take advantage of the immersive capabilities offered by VR to increase student learning outcomes. Using VR, a student could access and complete course topics at their own pace, proceeding through the course as dependencies are completed. This student-centered pedagogy can possibly increase students' engagement according to existing research work [12-15]. In addition, the efficacy of VR in improved knowledge retention and learning motivation could show a great deal of promise in fields with complex topics and topic relationships [16-17]. Engineering and computing are an ideal fit, as the interdisciplinary knowledge required to succeed in many courses benefits the topic and dependency-based structure of the student-centered learning pedagogy.

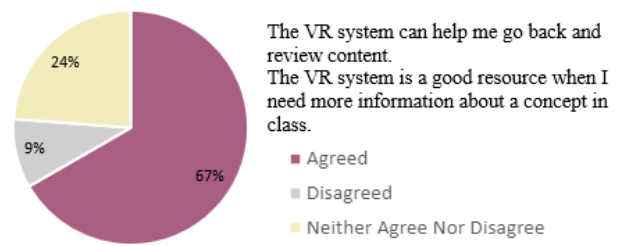

Fig. 5. This graph shows the results of two questions from the post-questionnaire. Though the allotment of responses to both questions was identical, many students responded differently to both questions.

## V. Conclusion and future work

In this paper, the major contributions are the proposal of a student-centered learning pedagogy utilizing VR, the development of an application to teach the oculomotor examination, and the user study conducted to collect feedback from students. Based on the results of the user study, with further development, there is potential for increased learning outcomes for courses that combine a student-centered learning framework with VR. The sample size of 21 for this study was small, and more data must be collected to validate these conclusions.

In the future, we plan to collect more user data and improve our VR application based on users' feedback. The application and study method will be adjusted to allow students to administer and take the test, and the study will be expanded to further analyze the impacts of student-centered learning when applied with VR.

## VI. Acknowledgment

This material is based in part upon work supported by: The National Science Foundation under grant number(s) NSF IUSE: EHR award # 2142428 and #1730568. Any opinions, findings, and conclusions or recommendations expressed in this material are those of the author(s) and do not necessarily reflect the views of the National Science Foundation.

## References

[1] P. C. Shill et al., "WIP: Development of a Student-Centered Personalized Learning Framework to Advance Undergraduate Robotics Education," unpublished.

[2] M. Coban, Y. I. Bolat, and I. Goksu, "The potential of immersive virtual reality to enhance learning: A meta-analysis," Educ. Res. Rev., vol. 36, p. 100452, 2022, doi: 10.1016/j.edurev.2022.100452.

[3] Q. Zhang, K. Wang, and S. Zhou, "Application and practice of vr virtual education platform in improving the quality and ability of college students," IEEE Access, vol. 8, pp. 162830–162837, 2020, doi: 10.1109/ACCESS.2020.3019262.

[4] Y. Zhong and J. J. Liu, "How Does the Virtual Reality work on Science Education: An Exploratory Study of Emotional Effects on Learning," in 2022 IEEE 2nd International Conference on Educational Technology, ICET 2022, 2022, pp. 105–109, doi: 10.1109/ICET55642.2022.9944496.

[5] V. Clay, P. König, and S. König, "Eye tracking in virtual reality," J. Eye Mov. Res., vol. 12, no. 1, 2019, doi: 10.16910/jemr.12.1.3.

[6] H. Kim, Y. T. Kwon, H. R. Lim, J. H. Kim, Y. S. Kim, and W. H. Yeo, "Recent Advances in Wearable Sensors and Integrated Functional Devices for Virtual and Augmented Reality Applications," Adv. Funct. Mater., vol. 31, no. 39, p. 2005692, 2021, doi: 10.1002/adfm.202005692.

[7] A. Leube, K. Rifai, and S. Wahl, "Sampling rate influences saccade detection in mobile eye tracking of a reading task," J. Eye Mov. Res., vol. 10, no. 3, 2017, doi: 10.16910/jemr.10.3.3.

[8] B. T. Hathiram and V. S. Khattar, "Videonystagmography," Int J Otorhinolaryngol Clin, vol. 4, no. 1, pp. 17–24, 2012.

[9] A. Dæhlen, I. Heldal, and Q. Ali, "Technologies Supporting Screening Oculomotor Problems: Challenges for Virtual Reality," 2023.

[10] R. Lindgren, M. Tscholl, S. Wang, and E. Johnson, "Enhancing learning and engagement through embodied interaction within a mixed reality simulation," Comput. Educ., vol. 95, pp. 174–187, 2016, doi: 10.1016/j.compedu.2016.01.001.

[11] J. Dinet and M. Kitajima, "Immersive interfaces for engagement and learning: Cognitive implications," ACM Int. Conf. Proceeding Ser., vol. 323, no. 5910, pp. 66–69, 2018, doi: 10.1145/3234253.3234301.

[12] E. L. Deci and R. M. Ryan, "Self-determination theory.," 2012.

[13] G. R. Rasmussen, "An evaluation of a student-centered and instructor-centered method of conducting a graduate course in education," J. Educ. Psychol., vol. 47, no. 8, pp. 449–461, 1956, doi: 10.1037/h0043689.

[14] C. L. Hovey, L. Barker, and M. Luebs, "Frequency of instructor- And student-centered teaching practices in introductory CS courses," in SIGCSE 2019 - Proceedings of the 50th ACM Technical Symposium on Computer Science Education, 2019, pp. 599–605, doi: 10.1145/3287324.3287363.

[15] S. Pedersen and M. Liu, "Teachers' beliefs about issues in the implementation of a student-centered learning environment," Educ. Technol. Res. Dev., vol. 51, no. 2, pp. 57–76, 2003, doi: 10.1007/BF02504526.

[16] J. K.-Y. Essoe, N. Reggente, A. A. Ohno, Y. H. Baek, J. Dell'Italia, and J. Rissman, "Enhancing learning and retention with distinctive virtual reality environments and mental context reinstatement," npj Sci. Learn., vol. 7, no. 1, p. 31, 2022.

[17] C. Ekstrand, A. Jamal, R. Nguyen, A. Kudryk, J. Mann, and I. Mendez, "Immersive and interactive virtual reality to improve learning and retention of neuroanatomy in medical students: a randomized controlled study," Can. Med. Assoc. Open Access J., vol. 6, no. 1, pp. E103--E109, 2018.